# Thomas-Wigner rotation and Thomas precession: actualized approach

Alexander L. Kholmetskii[1)] and Tolga Yarman[2)]

[1]Department of Physics, Belarusian State University, 4 Nezavisimosti Avenue 220030, Minsk, Belarus, E-mail: khol123@yahoo.com
[2]Okan University, Akfirat, Istanbul, Turkey

**Abstract:** We show that the explanation of Thomas-Wigner rotation (TWR) and Thomas precession (TP) in the framework of special theory of relativity (STR) contains a number of points of inconsistency, in particular, with respect to physical interpretation of the Einstein velocity composition law in successive space-time transformations. In addition, we show that the common interpretation of TP falls into conflict with the causality principle. In order to eliminate such a conflict, we suggest considering the velocity parameter, entering into expression for the frequency of TP, as being always related to a rotation-free Lorentz transformation. Such an assumption (which actually resolves any causal paradoxes with respect to TP), comes however to be in contradiction with the *spirit* of STR. The results obtained are discussed.

## 1. Introduction

Since the development of special theory of relativity (STR), numerous publications were devoted to the Thomas-Wigner rotation (TWR) and Thomas precession (TP), which applied either analytical (e.g. [1-6]) or geometrical (e.g. [7-9]) approaches. There is a common agreement that both of these effects, being closely related to each other, have a purely kinematical origin, which thus signifies that their physical mechanism, in fact, remains unclarified. In this respect, TWR and TP are unique, because any other implication of STR can be understood at the model level with the involvement of basic properties of empty space-time postulated by STR, and the established measurement procedures.

Thus, the recognition of purely kinematical origin of TWR and TP reflects, in our opinion, the neglect of the problem to look for the physical meaning of these effects, which seems especially unsatisfactory in the situation, where, as known, TP affects the dynamical properties of moving objects, in particular, in the semi-classical explanation of spin-orbit coupling in the atomic physics [1, 3]. If so, one believes that the real physical mechanism, explaining TP, should exist. Here we mention that the attempts to relate TP with the time dilation effect in a rotating object are formally correct (see e.g. [10]), but at the same time, do not add lucidity to the clarification of TP mechanism. We can add that this ambiguous situation gave rise to a number of paradoxes with respect to TP (e.g., [11-15]), where the authors, in fact, develop the feeling that the entire physical context of STR should be revisited. At the same time, a major part of these paradoxes finds a non-contradictory relativistic explanation (see, e.g. [10]). However, there exist a paradox by Bacry [15], which, in our opinion, did not still find its satisfactory resolution, and in the present contribution we will separately consider this paradox and its generalized version in sub-section 2.2.

Further, one has to remember that TP represents the particular case of TWR, and any attempt to disclose the physical mechanism of TP is directly related to a search of physical meaning of TWR.

As known, TWR emerges in successive space-time transformations between three inertial reference frames K, K′ and K″, moving with the relative velocities *v* and *u*, correspondingly, which are not collinear to each other [4, 5]. In this case, applying rotation-free Lorentz transformations between K to K′, and then from K′ to K″, we obtain that the systems K and K″ are not parallel to each other, and their axes are turned out with respect to each other at some angle to be found at the first time by Thomas [1] and Wigner [4]. This effect, being applied to a point-like particle with a designated axis (e.g. spin), moving along a curved path, causes the precession of its spin (TP), which, in particular, reduces twice the spin-orbit interval in atoms [1]. We remind that this result can be directly derived from the Dirac equation for a bound electron (see e.g. [16]), without addressing to the classical analogy.

The goal of the present paper is to look closer at the physical mechanisms of TWR and TP on the basis of known relativistic effects. This way we, first of all, address to the definition of rotation-free Lorentz transformation and point out that it does not imply, in general, the hold up of the parallelism of coordinate axes of the involved inertial reference frames. It is surprising that this important circumstance is not usually mentioned in the analysis of successive space-time transformations. Therefore, for a better understanding of TWR, below we suggest determining not only the rotational angles between such frames, but also to specify separately the spatial orientation of their axes, as viewed by different inertial observers. In order to simplify the solution of this problem, we apply a convenient methodological trick. That is, instead of the implementation of space-time transformations between various inertial frames (which further demands to take into account the relativity of simultaneity of events, when the spatial directions of coordinate axes are determined for different observers), we directly operate with vectorial quantities, whose relativistic transformations are well defined (e.g., the electric/magnetic fields, electric/magnetic dipoles, etc.; in the present paper we use the electric dipole moment of a point-like particle). Thus, if this vector coincides, say, with the direction of the axis *x*′ in its rest frame K′, then the transformation of this vector to another frame K directly defines the direction of the axis *x*′ for an observer in K.

For further simplification we consider the motion of inertial frames in the *xy* plane only, and in the analysis of TWR we deal with successive Lorentz transformations with orthogonal relative velocities, at least between two frames. Besides, the major part of our calculations is carried out to the accuracy $c^{-2}$, where *c* is the light velocity in vacuum. These assumptions make the mathematical side of our analysis to be very simple, which allows us to focus our attention to its physical interpretation. This way we derive, first of all, the known expression for TWR in its application to our particular problems. We also show that TWR can be derived via considering, for example, the transformation of electric dipole moment (as the representative of convenient vectorial quantity mentioned above), which additionally allows us to determine directly the spatial orientation of coordinate axes of involved systems for different inertial observers. This way we specify the notions of parallel and non-parallel coordinate systems (section 2) and point out that such coordinate systems, in general, are not necessarily Cartesian, as viewed by different observers. This result is in odd with the often-used interpretation of TWR as a simple rotation of coordinate axes of two inertial Cartesian systems and shows serious inconsistencies in the common kinematical explanations of TWR (sub-section 2.1).

Analyzing further TP as the particular representative of TWR in sub-section 2.2, we emphasize the essential difference between these effects, in spite of their common fundamental origin. Namely, in TWR we deal with inertial frames, belonging to different inertial observers, where the relationship between their relative velocities, in general, might be arbitrary. In contrast, TP emerges in the successive space-time transformations between Lorentz frames, co-moving with the *same particle*, propagating along a curved path. Therefore, the corresponding relative velocities of these frames are in a causal relation between each other, when the velocity and acceleration of particle are known. In the particular case of a circular motion of particle, we show that the rest frames of moving particle, taken at the time moments *t* and (*t+dt*), do experience TWR, remaining to be Cartesian for observers in these frames. However, this is not, in gen-eral, the case, when the particle at hand moves along a curved path of an arbitrary smooth shape, and the revealed inconsistencies found with respect to TWR, also emerge in the case of TP. These inconsistencies are well manifested in the mentioned above paradox by Bacry [15] and its "inverse" version, presented in sub-section 2.2. Finally, we conclude in section 3.

## 2. Successive space-time transformations and relativistic analysis of Thomas-Wigner rotation and Thomas precession

In this section we analyze TWR (subsection 2.1) and TP (sub-section 2.2) on the basis of well known properties of the Lorentz group and disclose a number of points of inconsistency in their common interpretation.

First of all, let us remind the definition of parallel coordinate systems K and K′ given, for example, in ref. [17]. Namely, we define K and K′ to be parallel to each other, if their relative velocities have equal but opposite projections on their respective coordinate axes. In other words, if an observer in K sees the velocity of K′ to be equal to $V$, then an observer in K′ sees the velocity of K to be equal to $-V$. Correspondingly, the Lorentz transformations $\mathbf{L}(V)$ and $\mathbf{L}(-V) = \mathbf{L}^{-1}(V)$ are both rotation-free.

At the same time, this definition does not imply that an observer in K sees the coordinate system of K′ to be Cartesian, and vice versa. Such a situation is depicted in Fig. 1, where for an observer in K (Fig. 1a) the axes $x'$ and $y'$ of the frame K′ do not constitute the angle $\pi/2$. The symmetric situation with respect to an observer in K′ is shown in Fig. 1b. The change of spatial orientation of coordinate axes of a moving system in comparison with the case $V=0$ is caused by the contraction of scale along the direction of $V$. Indeed, for an observer in K (see Fig. 1a), any designated segment $X'$ belonging to the axis $x'$ of K′ has the unchanged projection onto the direction, orthogonal to $V$; however, its projection along $V$ contracts by $1/\gamma$ times, where $\gamma = (1-V^2/c^2)^{-1/2}$ is the Lorentz factor. As a result, the entire axis $x'$ is rotated at some angle $\vartheta_x$ with respect to the axis $x$, and the value of $\vartheta_x$ will be found below. Just the same effect induces the rotation of the axis $y'$ with respect to the axis $y$ at the angle $\vartheta_y$ having the opposite sign in comparison with the rotational angle between the axes $x'$ and $x$. Hence the angle between axes $x'$ and $y'$, as seen in K, is equal to $(\pi/2 + \vartheta_x + \vartheta_y)$, though one can show that the systems K and K′ are parallel to each other, according to the definition presented above. Indeed, an observer in K sees the projections of velocity of K′ on the axes $x$ and $y$ to be equal to $V\cos\alpha$, $V\sin\alpha$, correspondingly (Fig. 1a), while an observer in K′ fixes the projections of velocity of K on the axes $x'$ and $y'$ to be respectively equal to $-V\cos\alpha$, $-V\sin\alpha$ (Fig. 1b), and the definition of parallel coordinate systems is perfectly fulfilled with respect to K and K′.

In order to determine the angles $\vartheta_x$, $\vartheta_y$, we use the relativistic transformation of length in the rotation-free Lorentz transformation [5]:

$$\mathbf{r} = \mathbf{r}' - \frac{(\gamma-1)}{\gamma V^2}(\mathbf{r}' \cdot \mathbf{V})\mathbf{V}. \tag{1}$$

The angle $\vartheta_x$ between the axes $x$ and $x'$ can be found, when we choose $\mathbf{r}'$ with the projections $\{X',0\}$ in the frame K′. Hence

$$X_x = X' - \frac{(\gamma-1)}{\gamma V^2}(X'V\cos\alpha)V\cos\alpha = \frac{X'}{\gamma}(\gamma\sin^2\alpha + \cos^2\alpha), \tag{2a}$$

$$X_y = -\frac{(\gamma-1)}{\gamma V^2}(X'V\cos\alpha)V\sin\alpha = -\frac{(\gamma-1)X'}{\gamma}\sin\alpha\cos\alpha, \tag{2b}$$

and

$$\mathrm{tg}\,\vartheta_x = \frac{X_y}{X_x} = -\frac{(\gamma-1)\sin\alpha\cos\alpha}{\gamma\sin^2\alpha + \cos^2\alpha}. \tag{3}$$

To the accuracy of calculations $c^{-2}$, eq. (3) yields:

$$\vartheta_x \approx -\frac{V^2}{2c^2}\sin\alpha\cos\alpha = -\frac{V_x V_y}{2c^2}, \qquad (4)$$

where we adopted the sign "*minus*" for the *clockwise* rotation.

In a similar way one can show that the angle between the axes $y$ and $y'$ is defined by the equation

$$\text{tg}\,\vartheta_y = \frac{(\gamma-1)\sin\alpha\cos\alpha}{\sin^2\alpha + \gamma\cos^2\alpha}, \text{ and}$$

$$\vartheta_y \approx \frac{V^2}{2c^2}\sin\alpha\cos\alpha = \frac{V_x V_y}{2c^2}, \qquad (5)$$

which in the adopted accuracy of calculations has the same magnitude, as that of $\vartheta_x$, but the reverse sign (i.e. the counter clockwise rotation).

Eqs. (4), (5) show that the angles $\vartheta_x$, $\vartheta_y$ are both vanishing, when one of the projections of $V$ (either $V_x$, or $V_y$) is equal to zero. It corresponds to the situation, when the relative velocity lies along of one of coordinate axes (i.e., the special Lorentz transformation). In this case the coordinate axes of the systems K and K' keep their parallelism and remain Cartesian for each other.

Having obtained these results, we further carry out the analysis of TWR (sub-section 2.1) and TP (sub-section 2.2).

### 2.1. TWR in successive space-time transformations with orthogonal relative velocities.

Now we consider the successive space-time transformations from the frame K″ to K′, and then from the frame K′ to K, when the velocity of K″ in K′ is equal to $u\{0, u, 0\}$, and the velocity of K′ in K is equal to $v\{v, 0, 0\}$ (see Fig. 2). In the succession of these transformations we will determine the directions of coordinate axes of the frame K″ with respect to the corresponding axes of the frame K via applying the relativistic transformations for electric $p$ and magnetic $m$ dipole moments of a small electrically neutral particle [18]:

$$\boldsymbol{p} = \boldsymbol{p}_0 - \frac{(\gamma-1)}{v^2}(\boldsymbol{p}_0\cdot\boldsymbol{v})\boldsymbol{v} + \frac{\boldsymbol{v}\times\boldsymbol{m}_0}{c}, \qquad (6)$$

$$\boldsymbol{m} = \boldsymbol{m}_0 - \frac{(\gamma-1)}{v^2}(\boldsymbol{m}_0\cdot\boldsymbol{v})\boldsymbol{v} + \frac{\boldsymbol{p}_0\times\boldsymbol{v}}{c}. \qquad (7)$$

Here $\boldsymbol{p}_0$ ($\boldsymbol{m}_0$) is the proper electric (magnetic) dipole moment measured in its rest frame.

Further we assume that $m_0=0$, and the dipole is at rest in the frame K″.

First consider the case, where the electric dipole moment is parallel to the axis $y''$, i.e. it has the projections $\{0, p'', 0\}$. Then for the frame K' the transformations (6), (7) yield

$$p'_x = p'_z = 0, \quad p'_y = \frac{p''}{\gamma}, \qquad (8a)$$

$$m'_x = m'_y = m'_z = 0. \qquad (8b)$$

The transformation from K' to K gives:

$$p_x = p_z = 0, \quad p_y = p'_y = \frac{p''}{\gamma}. \qquad (9)$$

Eqs. (9) show that in the succession of transformations K″→K′→K, the axes $y$ and $y''$ remain parallel to each other.

Next consider the situation, when the electric dipole moment is parallel to the axis $x''$, i.e. $\boldsymbol{p}_0\{p'', 0, 0\}$. Then for the frame K' the transformations (6), (7) yield

$$p'_x = p'', \quad p'_y = p'_z = 0, \qquad (10a)$$

$$m'_z = \frac{up''}{c}, \quad m'_x = m'_y = 0. \qquad (10b)$$

Thus, for an observer in the frame K we obtain:

$$p_x = \frac{p'_x}{\gamma} = \frac{p''}{\gamma}, \quad p_y = -\frac{vm'_z}{c} = -\frac{uvp''}{c^2}. \tag{11}$$

This equation shows that in the considered successive transformations the axes $x$ and $x''$ are no longer parallel to each other, but drow the angle $\varphi$ defined by the equality

$$tg\varphi = \frac{p_y}{p_x} = -\gamma\frac{uv}{c^2}. \tag{12}$$

To the accuracy of calculations $c^{-2}$, we have

$$\varphi \approx -uv/c^2. \tag{13}$$

Since the axes $y$ and $y''$ remain parallel to each other (see eq. (9)), we conclude that for an observer in K the coordinate system of K″ is not Cartesian, and the axes $x''$ and $y''$ make the angle ($\pi/2+\varphi$) with respect to each other.

Now let us carry out a direct rotation-free Lorentz transformation from an inertial frame K‴ to K with the relative velocity $\boldsymbol{V}\{v, u\sqrt{1-v^2/c^2}, 0\}$, where the presented components of $\boldsymbol{V}$ are found via the Einstein law of velocity composition [5] in the considered succession of transformations K″→K′→K. Then, according to Wigner [4], in this transformation the inertial frame K‴ (which, by definition, is parallel to K) differs from K″ by the Thomas-Wigner rotational angle Ω. In order to determine Ω, we first observe that the rotation-free transformation from K‴ to K reproduces the situation to be shown in Fig. 1a. Therefore, to the adopted accuracy of calculations $c^{-2}$, we can directly apply eqs. (4) and (5). Hence, for the indicated components of the velocity $\boldsymbol{V}$, we obtain the angle between the axes $x'''$ and $x$ to be equal to

$$\vartheta_x \approx -V_x V_y/2c^2 \approx -\frac{uv}{2c^2}, \tag{14}$$

and the angle between the axes $y'''$ and $y$

$$\vartheta_y \approx V_x V_y/2c^2 \approx uv/2c^2. \tag{15}$$

The directions of the axes $x'''$ and $y'''$ are shown in Fig. 2 by the thin arrows. Accordingly, we obtain the turn of the axis $x''$ with respect to $x'''$ at the angle

$$\Omega = \varphi - \vartheta_x = -uv/2c^2, \tag{16}$$

and the turn of the axis $y''$ with respect to $y'''$ at the same angle

$$\Omega = -\vartheta_y = -uv/2c^2, \tag{17}$$

as seen in the frame K. Thus we actually see the coordinate system of K″ to be turned as the whole with respect to coordinate system of K‴ at the Thomas-Wigner angle (17).

One can check that the general expression for the Thomas-Wigner angle (see, e.g. [5]), being applied to the problem in Fig. 2, indeed yields the same result (17) to the adopted accuracy $c^{-2}$.

In accordance with this result, we observe that the frames K and K″ do not satisfy to the definition of parallel systems presented above. Indeed, the velocity $\boldsymbol{V}$ of K″ in K has the components $\{v, u\sqrt{1-v^2/c^2}, 0\}$, whereas the velocity $\boldsymbol{V''}$ of K in K″ has the components $\{-v\sqrt{1-u^2/c^2}, -u, 0\}$, which are straightforwardly found from the Einstein law of velocity composition. Thus $V_x \neq -V''_x$, $V_y \neq -V''_y$, and the definition of parallel systems is not fulfilled. Taking also into account that $V^2 = V''^2$ (which reflects the reciprocity principle [19]), we conclude that the frames K″ and K‴ differ from each other via a mere spatial rotation.

At the same time, now we especially emphasize that both frames, K″ and K‴, are not Cartesian for an observer in K, and this fact becomes crucial, when we consider the known attempts to prescribe a kinematical meaning to the TWR angle (17) on the basis of the Einstein law of velocity composition (see, e.g. [20]).

Such an attempt can be made, for example, via the comparison of components of $\mathbf{V}\{v, u\sqrt{1-v^2/c^2}, 0\}$ and $\mathbf{V}''\{-v\sqrt{1-u^2/c^2}, -u, 0\}$, assuming for simplicity that both values of $v$ and $u$ are positive. (This corresponds to the situation depicted in Fig. 2). Then, comparing the components of $\mathbf{V}$ and $\mathbf{V}''$, we conclude that an observer in K sees the magnitude of the $x$-component of velocity of K″ (equal to $v$) to be *largler* than the magnitude of the same component as seen in K″ ($\left|-v\sqrt{1-u^2/c^2}\right|$). At the positive $x$- and $y$-components of $\mathbf{V}$, it is possible only in the case, where the angle between the axis $x$ and $\mathbf{V}$ is *smaller*, than the angle between the axis $x''$ and $\mathbf{V}$, as seen in K. It means that in the frame K, the axis $x''$ is turned out with respect to the axis $x$ in the clock-wise direction (i.e. at the negative angle). Via the straightforward calculations one can show that this angle is equal to $\Omega = -uv/2c^2$.

Further on, an observer in K derives that the magnitude of $y$-component of $\mathbf{V}$ (to be equal to $u\sqrt{1-v^2/c^2}$) must be *smaller* than the magnitude of the same component on the axis $y''$ (to be equal to $|-u|$). Hence the angle between the axis $y$ and $\mathbf{V}$ is *larger*, than the angle between the axis $y''$ and $\mathbf{V}$, as seen in K. It means that for an observer in K the axis $y''$ is turned out with respect to the axis $y$ in the clock-wise direction, and numerically this negative angle is again equal to $\Omega = -uv/2c^2$. Thus the Einstein law of velocity composition requires the *common rotation* of the $x''$ and $y''$ axes *with respect* to $x$ and $y$ axes at the Thomas-Wigner angle $\Omega$ in the clock-wise direction, as seen in K. Since an observer in K sees his own coordinate system to be Cartesian, the system of K″ frame *must* remain Cartesian in K, too.

However, we have seen above via the relativistic transformation of an electric dipole moment that this is not the case. Namely, the system of K″ is no longer Cartesian in K: in particular, the axes $y$ and $y''$ remain parallel to each other, while the axes $x$ and $x''$ make the angle $\varphi$ defined by eq. (13), see again Fig. 2. Such orientation of the axes of K″ with respect to the axes of K makes impossible to explain the observed components of $\mathbf{V}$ via the Eientein law of velocity composition. In particular, now we have to conclude that $V_y = -V''_y$, which is obviously at odd with this law. The inequality $V_x \neq -V''_x$ holds, but with the angle (13) between the axes $x$ and $x''$ (which is twice of the Thomas-Wigner rotational angle $\Omega$), the relationship between these components becomes incompatible with the velocity composition law, too.

Thus, we find an inconsistency in the kinematical explanation of TWR. Namely, the non-contradictory explanation of the velocity composition law in the successive space-time transformations requires the common Thomas-Wigner rotation of the axes $x''$, $y''$ *with respect to the axes $x$, $y$*, whereas our analysis shows that the axes $x''$, $y''$ experience TWR *with respect to the axes $x'''$, $y'''$* (but not with respect to $x$, $y$ axes, see Fig. 2).

As we are aware, *this inconsistency has not pinned down to the moment.*

### 2.2. Thomas precession at a circular motion

We again consider some simple situations, which make elementary the required calculations, allowing us to concentrate attention to the physical meaning of the obtained solutions.

Namely, let us consider the case, where a particle with spin is moving on a circular orbit, when its tangential velocity $\mathbf{v}$ and acceleration $\dot{\mathbf{v}}$ are orthogonal to each other. In addition, like in the analysis of TWR, we restrict our calculations by the accuracy $c^{-2}$. In this case the frequency of TP takes the form [10]

$$\omega_T = \frac{(\gamma-1)}{\gamma}\frac{|\mathbf{v}\times\dot{\mathbf{v}}|}{v^2} \approx \frac{v\dot{v}}{2c^2}, \qquad (18)$$

and we wish to understand the physical meaning of eq. (18).

We again should like to stress that in contrast to TWR (where the relationship between relative velocities in the successive space-time transformations can be taken arbitrarily), in the case of TP we deal with the succession of space-time transformations between Lorentz frames, comoving with the same particle at different time moments. Thus, at two time moments $t$ and $t+dt$, the relationship between the respective velocities of co-moving frames, $v$ and ($v + \dot{v}dt$), are fully determined by the kinematical characteristics of the particle's motion. This fact makes the corresponding angle of TWR $\Omega$ to be dependent on the velocity of particle and its acceleration, so that the Thomas rotation frequency $\omega_T = d\Omega/dt$ also depends on $v$ and $\dot{v}$, as eq. (18) indicates.

Further, directing the axis $x$ along the vector $v$ at the considered time moment $t$, we observe that in the case of a circular motion (when $v \perp \dot{v}$), the velocity of co-moving Lorentz frame at the moment $t+dt$ differs from $v$ by the component $\dot{v}dt$ directed along the axis $y$, see Fig. 3.

Thus, like in the analysis of TWR, we again deal with three inertial reference frames K (chosen as the laboratory frame), K′, K″, and this time the rotation-free Lorentz transformations are carried out between the frames K′, K (with the velocity $v$) and K″, K (with the velocity $v + \dot{v}dt$). We want to determine the mutual orientation of coordinate axes of K″ in K′.

First we observe that due to the adopted parallelism of $v$ and axis $x$, we have a special Lorentz transformation between K and K′, and the axes of these frames remain parallel to each other.

Further we notice that the motion of the frame K″ in K corresponds to the situation shown in Fig. 1a, with the components of velocity $V\{v, \dot{v}dt, 0\}$. Hence, due to the scale contraction effect, the axes $x''$ and $x$ make the angle (4), which in our case is equal to

$$\vartheta_x \approx -\frac{V_x V_y}{2c^2} = -\frac{v\dot{v}dt}{2c^2} \quad (19)$$

to the accuracy of calculations $c^{-2}$.

The angle between $y$ and $y''$ is defined by eq. (5) and equal to

$$\vartheta_y \approx \frac{V_x V_y}{2c^2} = \frac{v\dot{v}dt}{2c^2}. \quad (20)$$

In order to determine the directions of axes of K″ for an observer in K′, we again address to the transformation for electric/magnetic dipole moment (6), (7) in the succession of transformations K″→K→ K′, assuming that the dipole is at rest in K″.

First we want to find the orientation of the axis $x''$ with respect to $x$, choosing the electric dipole moment $p_0\{p'', 0, 0\}$. Hence we obtain in the transformation from K″ to K:

$$p_x = p'' - \frac{(\gamma-1)}{\gamma V^2}(p''V_x)V_x \approx p''\left(1 - \frac{v^2}{2c^2}\right), \quad (21)$$

$$p_y = -\frac{(\gamma-1)}{\gamma V^2}(p''V_x)V_y \approx -p''\frac{v\dot{v}dt}{2c^2}, \quad (22)$$

$$m_z = \frac{p''\dot{v}dt}{c}. \quad (23)$$

We see that, as expected, the angle between the axis $x''$ and $x$, $\vartheta_x \approx p_y/p_x$, is again given by eq. (19) to the adopted accuracy of calculations $c^{-2}$.

Now we carry out the transformation from K to K′ at their relative velocity $\{-v, 0, 0\}$:

$$p'_x \approx p_x\left(1 - \frac{v^2}{2c^2}\right) \approx p''\left(1 - \frac{v^2}{c^2}\right), \quad (24)$$

$$p'_y = p_y + \frac{vm_z}{c} \approx -p''\frac{v\dot{v}dt}{2c^2} + p''\frac{v\dot{v}dt}{c^2} = p''\frac{v\dot{v}dt}{2c^2}. \quad (25)$$

Hence we derive the angle between $x''$ and $x'$, as seen in K′, to be equal to

$$\Omega \approx \frac{p'_y}{p'_x} = \frac{v\dot{v}dt}{2c^2}. \quad (26)$$

Comparing eqs. (19) and (26), we observe that in the frame K′ the angle (26) between $x''$ and $x'$ changes its sign in comparison with the angle between $x''$ and $x$ (as seen in K), and becomes *positive* (counter clock-wise rotation).

Such a change of spatial orientation of the axis $x''$ in K′ in comparison with its spatial orientation in K can be understood via the relativity of simultaneity of events, happening on the axis $x$ ($x'$) of K (K′). Indeed, let us designate the segment $X''$ on the axis $x''$ and fix in the frame K the time moment $t=0$, when the left end of this segment passes the axis $x$. Since an observer in K sees the axis $x''$ to be inclined with respect to the axis $x$ at the *negative* angle $\vartheta_x$ (eq. (19)), the right end of the segment $X''$ will pass the axis $x$ at the *later time moment* $\delta t \approx X''|\vartheta_x|/\dot{v}dt = X''v/2c^2$ (to the accuracy $c^{-2}$). However, due to the relativity of simultaneity of events in K and K′ frames, for an observer in K′ this event happens at the *earlier time moment* $\delta t' \approx X''v/2c^2 - X''v/c^2 = -X''v/2c^2$ in comparison with the event at $t'=0$, when the left end of the segment $X''$ passes the axis $x'$. It means that that the segment $X''$ (and thus the entire axis $x''$) is seen in K′ to be turned at the *positive angle* $\Omega$ with respect to the axis $x'$, which is confirmed by eq. (26) derived via the relativistic transformation of an electric dipole moment.

Finally, let us calculate the angle between the axes $y''$ and $y'$ for an observer in the frame K′, choosing $\boldsymbol{p}_0\{0, p'', 0\}$. First of all, using eqs. (6), (7) we obtain in the transformation from K″ to K:

$$p_x = -\frac{(\gamma-1)}{\gamma V^2}(p''V_y)V_x \approx -p''\frac{v\dot{v}dt}{2c^2}, \tag{27}$$

$$p_y = p'' - \frac{(\gamma-1)}{\gamma V^2}(p''V_y)V_y \approx p''\left(1 - \frac{\dot{v}^2(dt)^2}{2c^2}\right) = p'' \tag{28}$$

$$m_z = -\frac{p''V_x}{c} = -\frac{p''v}{c}. \tag{29}$$

Next we implement the transformation from K to K′ at the velocity $\{-v, 0, 0\}$:

$$p'_x = p_x - \frac{(\gamma-1)}{\gamma v^2}(p_x v)v \approx -p''\frac{v\dot{v}dt}{2c^2}, \tag{30}$$

$$p'_y = p_y + \frac{vm_z}{c} \approx p''\left(1 - \frac{v^2}{c^2}\right). \tag{31}$$

Here we notice that the appearance of negative projection $p'_x$ (eq. (30)) means the rotation of $y$-component of electric dipole moment in the counter clock-wise direction (adopted by us as positive) for an observer in K′. Hence the angle between the axes $y''$ and $y'$ is positive and equal to

$$\Omega \approx \frac{|p'_x|}{p'_y} \approx \frac{v\dot{v}dt}{2c^2}, \tag{32}$$

which is equal to the angle between axes $y''$ and $y$ (see eq. (20)).

We stress that the same spatial orientation of the axis $y''$ in the frames K and K′ is explained by the fact that the motion of K′ along the axis $x$ of K does not affects the simultaneity of events happening on the axis $y$ ($y'$).

Further we point out that the angle (26) between the axes $x''$ and $x'$ coincides with the angle (32) between the axes $y''$ and $y'$, and an observer in K′ sees the *common turn* of the system K″ in the counter-clock-wise direction at the Thomas-Wigner angle (32) (or (26)). Thus the frame K″ remains *Cartesian* in K′. Just this fact allows interpreting the TP as a real effect, which can be measured in experiments via the precession frequency

$$\omega_T = d\Omega/dt \approx v\dot{v}/2c^2$$

for the motion diagram of Fig. 3 (corresponding to the circular motion of particle).

In the case of motion of particle along a smooth curved path of an arbitrary shape, the vectors $\mathbf{v}$ and $\dot{\mathbf{v}}dt$, in general, are not orthogonal to each other, and the frame K″ is no longer Cartesian for an observer in K′. However, it is essential that the frequency of TP (18) is always determined by the component of $\dot{\mathbf{v}}dt$, which is orthogonal to $\mathbf{v}$. In these conditions the presence of a non-vanishing component of $\dot{\mathbf{v}}dt$ on the vector $\mathbf{v}$ (which makes the angle between the axes $x''$ and $y''$ different from $\pi/2$ for an observer in K′) occurs not essential in the calculation of $\omega_T$.

As known, the most convincing evidence of a real character of TP is done via the BMT equation [21] and its numerous practical applications, and now we would like to seek a physical explanation of TP, addressing to the motion diagram of Fig. 3. In particular, comparing the directions of the axes $x''$ and $y''$, as viewed in the frames K and K′, we see that in both of these frames the axis $y''$ has the same orientation, being turned with respect to the axis $y$ (or $y'$) at the positive angle $\Omega = v\dot{v}dt/2c^2$ (eq. (32)). However, the observers in K and K′ disagree with respect to the direction of the axis $x''$: in K it is turned in the clock-wise direction at the angle $-v\dot{v}/2c^2$ (eq. (19)), whereas in K′ it is turned in the counter clock-wise direction at the angle $+v\dot{v}/2c^2$ with respect to the axis $x$ ($x'$) (eq. (26)). From the physical viewpoint, this disagreement in the spatial orientations of the axis $x''$ for observers in K and K′, as explained above, is caused by the relativity of simultaneity of events that happened on the axis $x$ of K, and on the axis $x'$ of K′, which move at the velocity $v$ along the $x$-axis. This observation evokes the possibility to explain TP in terms of conventional relativistic effects of scale contraction and relativity of simultaneity of events. Indeed, an observer in K′ can ask an observer in K about the spatial orientation of the axes $x''$ and $y''$ in his frame, and to get answer that these axes have spatial orientations shown in Fig. 1a. Then an observer in K′ concludes that due to relativity of simultaneity of events in K and K′ frames, happening on the axis $x$ ($x'$), the direction of the axis $y''$ should be the same in both frames, while the axis $x''$, being in motion along the axis $y'$ of K′, should be turned at the positive angle $+v\dot{v}dt/c^2$ *with respect to its direction fixed in K*. If so, an observer in K′ finds the frame K″ to be Cartesian, turned as the whole at the angle $+v\dot{v}dt/2c^2$ relatively to his own frame.

The foregoing line of reasonings can be considered *as the physical explanation of TP*. However, it has a serious shortcoming. Indeed, in order to understand, why the system of K″ rotates with respect to the system of K′ at the angle $\Omega$, an observer in K′ has to ask observer in K about the spatial orientation of the axes $x''$ and $y''$, which he sees. Such a question obviously contradicts the *spirit* of STR, where all inertial frames are equivalent to each other and thus, any measurement procedures carried out by any observer in his own inertial frame, *must be sufficient* for understanding of any kinematical effects fixed in this frame.

As we have seen, this requirement is violated in the attempt to explain TP, presented above, though without the appearance of any intrinsic inconsistencies in the formal derivation of this effect, at least for a circular motion of particle with spin.

Even more serious problems in the analysis of TP emerge, when we consider the motion of a particle along a smooth path of arbitrary shape and look closer at eq. (18) defining the frequency of TP.

In particular, let us suppose that in some inertial reference frame K the precession frequency $\omega_T$ of spin of some moving particle is not equal to zero and hence, in a relativistically adequate situation, $\omega_T$ is not vanishing in any other inertial frame. At the same time, carrying out the transition from one inertial frame to another, we have to take into account that the relativistic transformations of velocity $\mathbf{v}$ and acceleration $\dot{\mathbf{v}}$ of the particle differ, in general, from each other (see, e.g. [5]). In these conditions we are not protected from the situation, where the cross product of velocity and acceleration is not zero in one inertial frame (i.e. $\omega_T \neq 0$), but becomes zero in another inertial frame (making $\omega_T = 0$), which comes into the obvious contradiction with causal requirements.

An example of such non-adequate situation is the paradox formulated by Bacry [15].

Let in some inertial reference frame K a particle with spin is moving along the axis *y* with the velocity $u(t)$ and acceleration $\dot{u}$, which are both parallel to this axis, see Fig. 4. Thus, $u(t) \times \dot{u} = 0$, and an observer in K sees no TP in his frame. However, an observer in another frame K′, moving against the axis *x* with the constant velocity *v*, measures the velocity of particle $V'\{v, u(t)\sqrt{1-v^2/c^2}, 0\}$ at any fixed time moment *t*, and the acceleration $\dot{u}'$, which, in general, is not collinear to $V'$. Therefore, in the frame K′ the cross product $(V' \times \dot{u}')$, in general, is not equal to zero, and an observer in this frame fixes the non-vanishing frequency of TP

$$\omega'_T \approx \frac{|V' \times \dot{u}'|}{2c^2}, \tag{33}$$

written to the accuracy $c^{-2}$. This result clearly contradicts the causality principle.

Malykin tries to resolve this paradox in the following way [10]: "The falcity of the paradox by Bacry can be proved as follows. .... The TP is explained by the relativity of concept of curvilinear translational motion of material points. If in one inertial reference frame S velocities of all point of the body at time *t* are the same, then in another inertial frame S ′, at time *t*′, they will be different at the curvilinear motion. The presence of the latter effect suggests that in SRT, in contrast to classical mechanics, there is no progressive curvilinear motion of extended body. Bacry's error lies in the fact that he is trying to ascribe a curved progressive motion to a solid body, observed in any frame of reference…" (This is the translation from Russian by the present authors).

However, this attempt to resolve the paradox by Bacry is obviously at odd with the definition of TP carefully done by Fisher [17]: "The Thomas precession is a kinematical effect which arises when studying *infinitesimally* extended objects. When extending these ideas to point particles with spin it shall be assumed that these particles…have infinitesimal spatial extent." Further, in accordance with this definition, Fisher formulates the postulate of "relativity-plus", expressed as follows [17]: "When an infinitesimal body experiences an infinitesimal change in velocity, the transfer of all elements to the final rest system is done simultaneously in both rest systems".

In the framework of this postulate, the Malykin explanation of the paradox by Bacry loses its force. We should add that Fisher also derives eq. (18) for the frequency of TP[1], so that the paradox by Bacry persists, and to the moment requires its resolution.

Moreover, we can suggest the "inverse" Bacry paradox, which is shown in Fig. 5, and which highlights the incorrectness of the available attempts to resolve the original Bacry paradox.

Let an observer in inertial frame K fixes the motion of particle in the plane *xy* with the velocity $V\{v, u(t), 0\}$, where the *x*-component *v* of this velocity is constant, while the *y*-component $u(t)$ changes with time[2]. Thus, the particle moves along a curved path, and its acceleration $\dot{u}$ lies along the axis *y*. Hence, according to eq. (18), the frequency of TP is equal to

$$\omega_T = \frac{(\gamma-1)}{\gamma} \frac{|V \times \dot{u}|}{V^2} \approx \frac{v\dot{u}}{2c^2} \tag{34}$$

(to the accuracy of calculations $c^{-2}$).

Next we introduce into consideration another inertial frame K′, which moves along the axis *x* of K at the constant velocity *v*. In this frame the particle has a single non-vanishing *y*-component of its velocity $u'(t) = u(t)\sqrt{1 - v^2/c^2}$, and the acceleration $\dot{u}'$ along the same axis. Therefore, to the adopted accuracy of calculations $c^{-2}$, in this frame

$$\omega'_T \approx u' \times \dot{u}'/2c^2 = 0, \qquad (35)$$

which obviously represents a non-adequate result from the relativistic viewpoint.

---

[1] The expression for the frequency of TP derived by Fisher differs from eq. (18) by the absence of factor $\gamma$ in the denominator. For the explanation of this difference, which is not essential in the analysis of Bacry paradox, see [10].

[2] The assumed constancy of *v* does not mean the absence of any force along the axis *x*. Indeed, in the absence of such force, the time derivative of *x*-component of momentum of particle with the mass *m* would be equal to zero, i.e. $d/dt(\gamma m v) = 0$. When the *y*-component of velocity changes with time, $d\gamma/dt \neq 0$, and hence $dv/dt \neq 0$, too. Therefore, the assumed condition *v=constant* implies the presence of a non-vanishing appropriate force component along the axis *x* of the frame K, which maintains the equality $dv/dt = 0$. We add that in the frame K′, which is introduced below, the *x*-component of total force, acting of the particle in this frame, disappears, and the acceleration $u'$ has the single *y*-component. We omit for brevity the proof of this statement, which is done via the relativistic transformation of force.

In our opinion, there is the only way to resolve both the original (Fig. 4) and "inverse" (Fig. 5) Bacry paradoxes. It is to demand that the velocity parameter *v* in eq. (18) always corresponds to rotation-free Lorentz transformation. Indeed, this constraint forbids to use the velocity *V′* in the calculation of TP in the original Bacry paradox (eq. (33)), because this velocity does not correspond to rotation-free Lorentz transformation (see Fig. 4). Considering further the inverse Bacry paradox, we see that the mentioned constraint makes eq. (35) non-applicable, since the velocity *u′* in this equation is not related to rotation-free transformation (see Fig. 5).

Thus, in the framework of this constraint, in the original Bacry paradox the absence of TP is fixed by any inertial observer, whereas in the inverse Bacry paradox the presence of TR is fixed by any inertial observer, too. Indeed, an arbitrary inertial observer K, looking at the motion of particle with spin along a curved path, is capable to calculate the spin precession frequency only after finding of another inertial frame K′, wherein the transformation between the rest frame of particle and K′ is rotation-free. At the next stage, the TP frequency is calculated according to eq. (18) (where *v* is understood as the velocity of this particle in K′), and then it is recalculated to the original frame K with taking into account the time dilation effect.

One can see that this algorithm makes absolute the fact of TP, which resolved any causal paradox.

At the same time, recovering relativistic causality in both problems of Fig. 4 and 5, we have to recognize that the requirement of rotation-free Lorentz transformation for the velocity *v* in eq. (18) comes in contradiction with the spirit of STR.

## 3. Conclusions

First of all, we stress a failure of STR to prescribe any real physical mechanisms to TWR and TP, and the common reference to "purely kinematical origin" of these effects, in our opinion, is not satisfactory.

Moreover, taking a closer look at TWR and TP, we reveal a number of points of inconsistency in their relativistic description:

- the impossibility to find the explanation for TWR via the notions of parallel and non-parallel systems and the Einstein velocity composition law. Namely, the obtained components of relative velocities onto the *x*- and *y*-axes, being found via the velocity composition law, occur at odd with the spatial orientations of these axes for corresponding inertial observers in three inertial reference frames K, K′ and K″, related by the successive Lorentz transformations;

- the impossibility to explain the mechanism of TP with the proper measurement tools in an inertial reference frame of observation, which is in contradiction with the basic STR postulate on the equivalence of all inertial frames;

- the persistence of the Bacry paradox, and the presentation of its "inverse" version, which makes the relativistic interpretation of TP incompatible with the causality principle.

In order to resolve these paradoxes, we suggest considering the velocity parameter, entering into expression for the frequency of TP (18), as being always related to a rotation-free Lorentz transformation. Such an assumption (which actually resolves any causal paradoxes with respect to TP), comes however into a contradiction with the spirit of STR, which is based on the postulate on the equivalence of all inertial reference frames.

The ways to overcome the mentioned above difficulties will be suggested in our subsequent paper on this subject, where the requirement of rotation-free Lorentz transformations for the velocity parameters of eq. (18) will be clarified with determination of its physical meaning.

# FIGURE CAPTIONS

Fig. 1. The relative motion of the frames K and K′ at the constant velocity $V$, constituting the angle $\alpha$ with the axis $x$. a – view from the frame K to the frame K′; b - view from the frame K′ to the frame K. The inertial systems K and K′ are parallel to each other according to the common definition, but the axes $x$, $x'$ and $y$, $y'$ are not parallel, and constitute the angle $\vartheta_x$ and $\vartheta_y$, correspondingly.

Fig. 2. Successive Lorentz transformations K″→K′→K with the orthogonal velocities $u$ (K″→K′) and $v$ (K′→K) lying in the $xy$-plane, in comparison with the direct rotation-free Lorentz transformation K‴→K with the velocity $V\{v, u\sqrt{1-v^2/c^2}, 0\}$. The directions of all axes of the frames involved are shown for an observer in K. (The axes of K, K′ and K″ are shown by the bold arrows. The axes of K‴ are shown by the thin arrows). We see that both systems, K‴ and K″, are not Cartesian for an observer in K, and K″ is turned out with respect to K‴ at the Thomas-Wigner angle $\Omega$ defined by eq. (17).

Fig. 3. Thomas precession for a circular motion of particle with spin, having the velocity $v$ at the time moment $t$, and the velocity $v + \dot{v}dt$ at the time moment $t+dt$ in the plane $xy$. The rotation-free Lorentz transformations are carried out between the frames K′, K at the velocity $v$, and between the frames K″, K at the velocity $v + \dot{v}dt$. The observers in the frames K and K′ agree with the direction of the axis $y''$ of the frame K″; however, due to relativity of simultaneity of evens happened on the axis $x$ (or $x'$), the observers in K and K′ disagree with respect to the direction of the axis $x''$ of K″. As a result, an observer in K sees the non-Cartesian system K″ with the angle between $x''$- and $y''$-axes $(\pi/2 + \vartheta_x + \vartheta_y)$, whereas an observer in K′ sees the system K″ to be Cartesian, turned out as the whole with respect to his frame at the Thomas-Wigner angle $\Omega$.

Fig. 4. The original paradox by Bacry. a – in the inertial frame K a particle moves along the axis $y$ with the constant acceletation, and its velocity $u(t)$ and acceleration $\dot{u}$ both lie along the axis $y$. Hence the Thomas precession of particle's spin is absent. b – in the inertial frame K′, moving agaist the axis $x$ of the frame K with the constant velocity $v$, the velocity of particle is equal to $V'\{v, u(t)\sqrt{1-v^2/c^2}, 0\}$, while its acceleration $\dot{u}'$ remains to be parallel to the axis $y$. Hence in this frame the cross product $(V' \times \dot{u}')$ is not equal to zero, and an observer in K′ has to fix the precession of spin of the particle.

Fig. 5. The "inverse" paradox by Bacry. a - In the inertial frame K a particle moves with the velocity $V\{v, u(t), 0\}$ lying in the plane $xy$, and with acceletation $\dot{u}$ along the axis $y$. The frequency of Thomas precession is proportional to the cross product $(V \times \dot{u})$ and is not equal to zero. b – In the inertial frame K′, moving with the constant velocity $v$ along the axis $x$ of the frame K, the velocity of particle $u'$ becomes parallel to the axis $y$, and the cross product of velocity and accelerating is vanishing. Thus, no Thomas precession is observed in the frame K′.

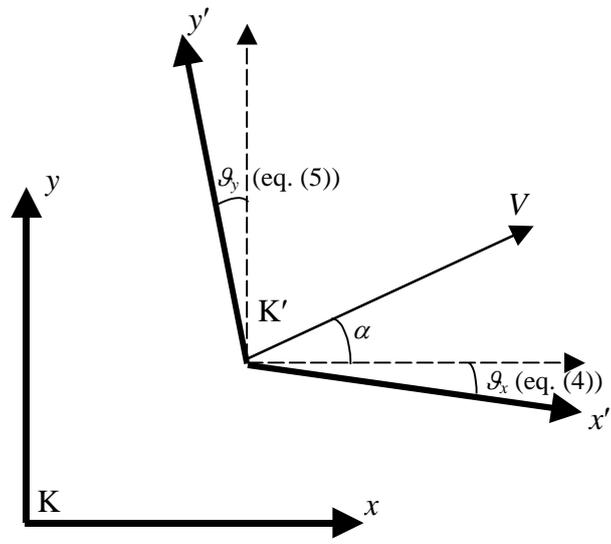

a)

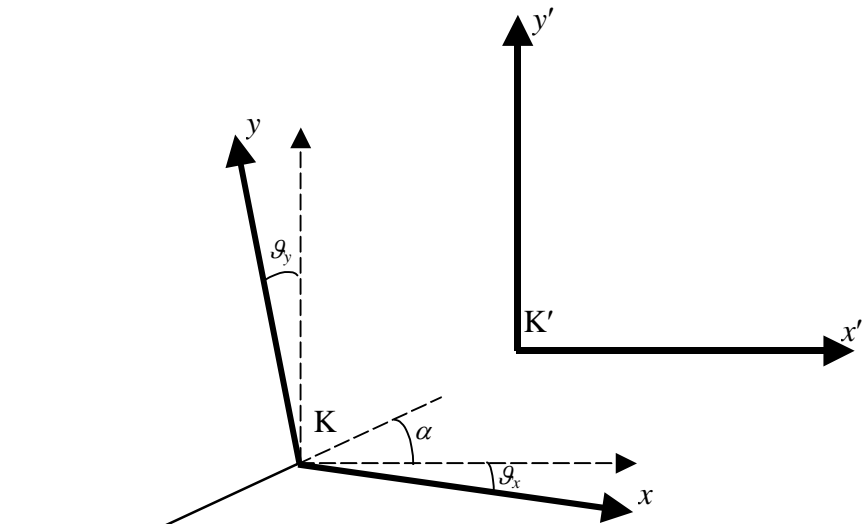

b)

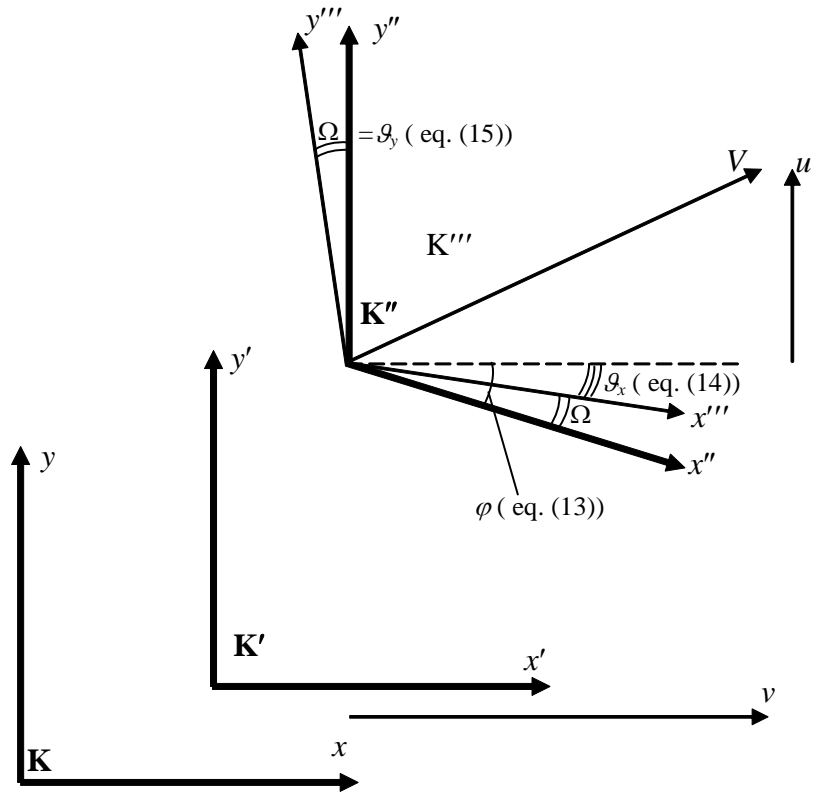

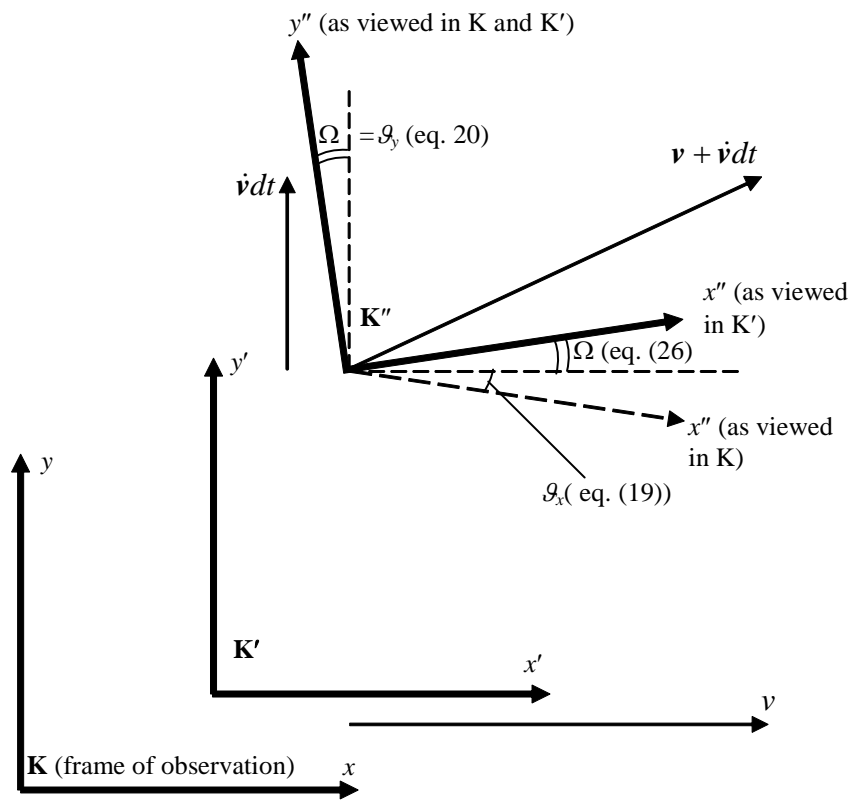

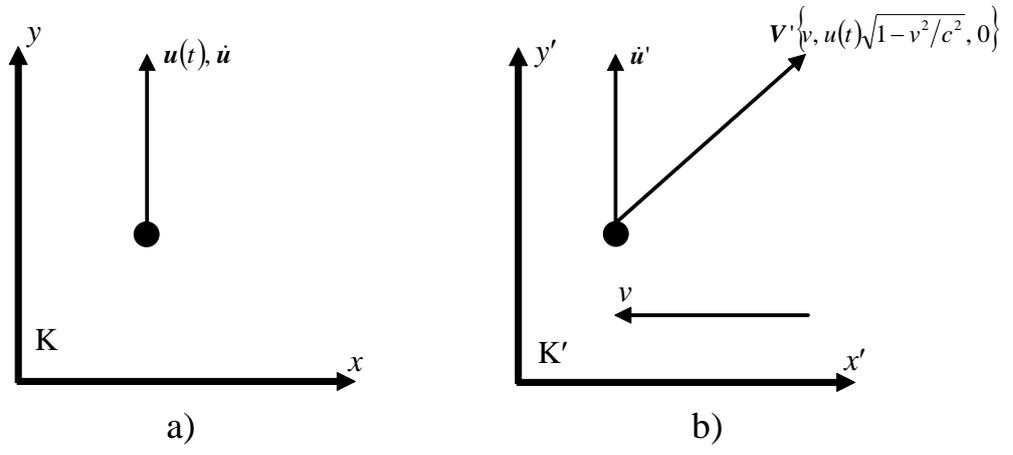

a)  b)

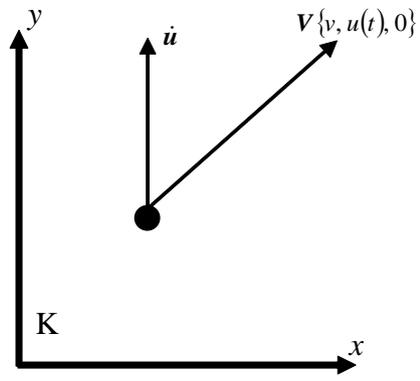 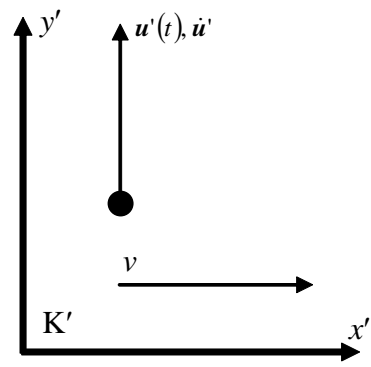

a)    b)